\documentclass[twocolumn]{aastex631}

\usepackage{times}
\usepackage{amsmath, amssymb}
\usepackage{url}
\usepackage{booktabs}

\usepackage[]{xcolor}
\definecolor{myg}{cmyk}{0.75002,0,1,0}

\definecolor{msnote}{hsb:rgb}{0.492,0.492,0.492}

\begin{document} 

\title{Covariance spectrum of MAXI J1820+070: On the nature of the Comptonizing flow}

\author{Shuai-Kang Yang}
\affiliation{Department of Astronomy, School of Physics and Technology, Wuhan University, Wuhan 430072, People’s Republic of China}

\correspondingauthor{Bei You}
\author{Bei You}
\affiliation{Department of Astronomy, School of Physics and Technology, Wuhan University, Wuhan 430072, People’s Republic of China}
\email{youbei@whu.edu.cn}

\author{Niek Bollemeijer}
\affiliation{Anton Pannekoek Institute for Astronomy, Amsterdam, Science Park 904, NL-1098 NH, The Netherlands}

\author{Phil Uttley}
\affiliation{Anton Pannekoek Institute for Astronomy, Amsterdam, Science Park 904, NL-1098 NH, The Netherlands}

\author{A.J. Tetarenko}
\affiliation{Department of Physics and Astronomy, University of Lethbridge, Lethbridge, Alberta, T1K 3M4, Canada}

\author{Andrzej A. Zdziarski}
\affiliation{Nicolaus Copernicus Astronomical Center, Polish Academy of Sciences, Bartycka 18, PL-00-716 Warszawa, Poland}

\author{Liang Chen}
\affiliation{Key Laboratory for Research in Galaxies and
Cosmology, Shanghai Astronomical Observatory, Chinese Academy of
Sciences, 80 Nandan Road, Shanghai 200030, China}

\author{P. Casella}
\affiliation{INAF-Osservatorio Astronomico di Roma, Via Frascati 33, I-00078 Monteporzio Catone, Italy}

\author{J.A. Paice}
\affiliation{School of Physics and Astronomy, University of Southampton, Southampton, SO17 1BJ, UK}

\author{Yang Bai}
\affiliation{Department of Physics, Fudan University, Shanghai 200438, People's Republic of China}

\author{Sai-En Xu}
\affiliation{Department of Astronomy, School of Physics and Technology, Wuhan University, Wuhan 430072, People’s Republic of China}

\begin{abstract}

We present an analysis of the covariance spectrum of the black hole X-ray binary MAXI J1820+070 during its hard state. For the first time, we extend coherence and covariance studies into the hard X-ray band up to $\sim$150 keV. We detect a clear drop in coherence above 30 keV on both short- and long-timescales relative to the 2–10 keV reference band. To investigate the origin of the coherent variability, we simultaneously fit the short- and long-timescale covariances and the time-averaged spectra with the Comptonization model. Surprisingly, the electron temperature associated with long-timescale variability is significantly higher than that on short timescales. Moreover, the temperature on long timescales remains relatively constant throughout the hard state, whereas the short-timescale temperature evolves with X-ray luminosity. We attribute the drop in coherence to multiple sources of seed photons, i.e., the blackbody and synchrotron photons. The independence between these two photon fields leads to the drop in coherence. Moreover, to explain the lower electron temperature on short timescales, we propose a two-Comptonization framework in which short-timescale variability arises from a vertically extended central region, while long-timescale variability originates at larger radii. The elevated geometry of the inner region leads to illumination primarily by cooler outer-disk photons, yielding a lower electron temperature. In this case, the evolution of the height of the elevated region could explain evolution of the electron temperature associated with the coherent variability throughout the hard state.

\end{abstract}

\section{Introduction} \label{sec:intro}

\begin{figure*}
\centering
\includegraphics[width=\textwidth]{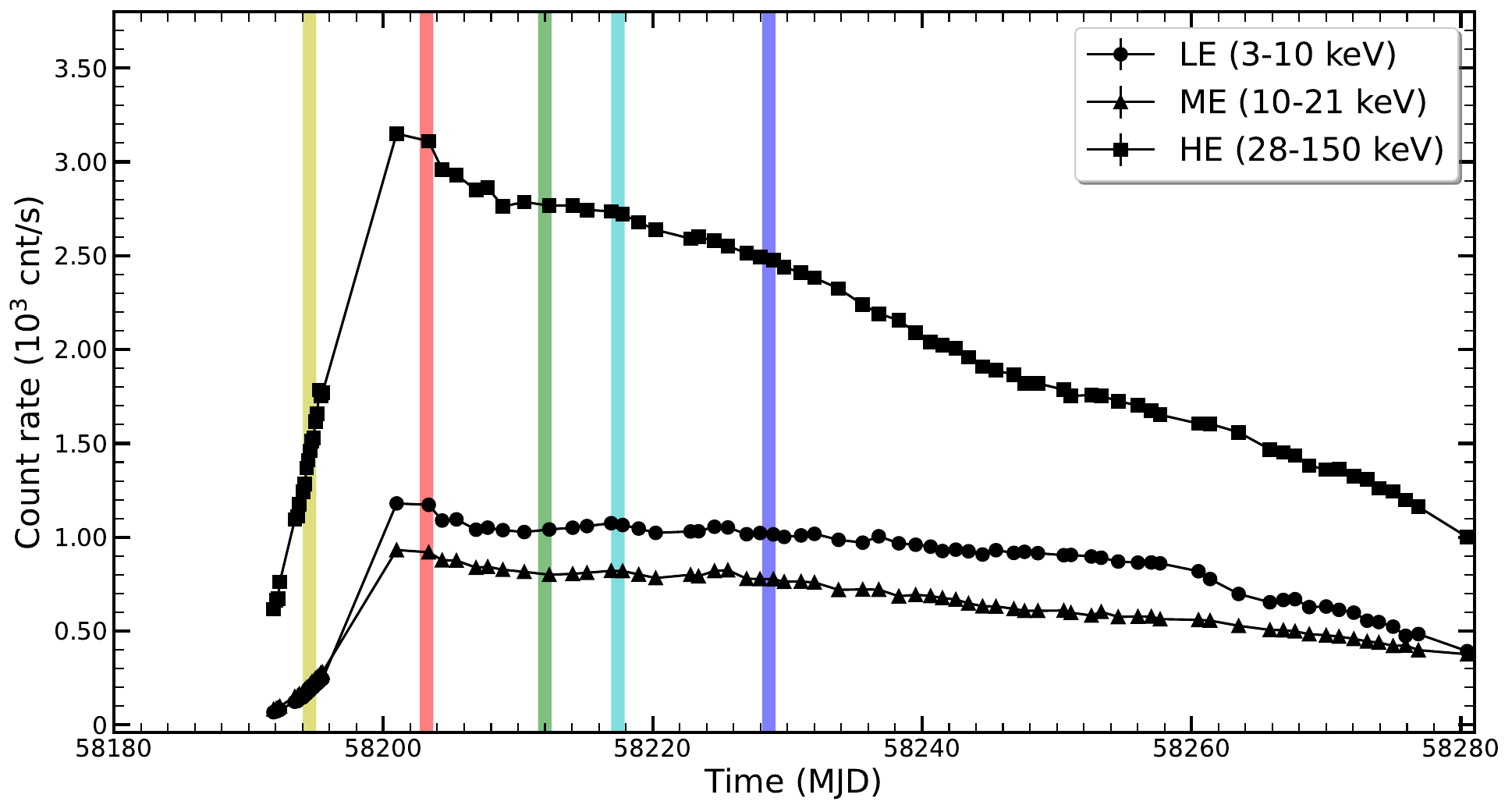}
\caption{
\emph{Insight}-HXMT light curves (in units of counts per second) of MAXI J1820+070 obtained from the HE (squares, 28–150 keV), ME (triangles, 10–30 keV), and LE (dots, 1–10 keV) detectors. Different colored shaded regions mark the five epochs selected for our spectral and timing analyses.
}
\label{lc}
\end{figure*}

In systems with accreting black holes (BHs), such as X-ray binaries (XRBs) that contain stellar-mass black holes and active galactic nuclei (AGNs) housing supermassive black holes, the observed X-ray photons primarily originate from the regions near the black holes\citep{Remillard2006,Done2007}. In the so-called hard state, the X-ray energy spectrum typically features two components: weak thermal emission in the low-energy band and strong power-law emission in the high-energy band \citep{Zdziarski2004,Belloni2010, Done2007}. The weak thermal emission is believed to originate from a geometrically thin, optically thick disk\citep{Shakura1973}.
The observed hard X-ray power-law continuum is commonly attributed to Comptonization of soft seed photons by hot electrons in the inner accretion flow\citep{Sunyaev1980,Paczynski1978,Svensson1994,Melia1993, Done2007}.

\subsection{Models of the accretion flow}

The advection-dominated accretion flow (ADAF), proposed by \cite{Narayan1994,Narayan1995}, has been widely used to explain the hard X-ray emission observed in BHXRBs, often referred to as a corona. This corona is commonly regarded as the primary region for Comptonization. The hot flow are postulated to exist within a specific radius, where the outer thin disk is truncated \citep{Esin1997,Done2007}. The co-evolution of the thin disk and the hot flow plays a crucial role in the state transitions observed during the outbursts of XRBs\citep{Esin1997, Belloni2010, Done2007}. \cite{Esin1997} proposed that the inner thin disk radius increases during the soft-to-hard state transitions. Recently, \cite{you2023} demonstrated that, in the framework of a truncated disk model, the truncation radius of MAXI J1820+070 increases as the mass accretion rate declines, indicating that the inner hot flow is expanding as this source transitions from the soft state to the hard state.
In this context, soft photons originating from the outer disk can act as seed photons. These photons are Compton up-scattered by thermal electrons within the hot flow through inverse Compton processes. This mechanism accounts for the observed power-law X-ray continuum. 

However, the nature of the Comptonizing hot flow is still a topic of debate. Alternative models have suggested that synchrotron emission may also account for the broad continuum features observed in XRBs \citep{Markoff2001,Markoff2005,Markoff&Sera2005}. And the synchrotron photons could also serve as the seed photons for the inverse-Compton scattering process \citep{Fabian1982,Markoff2005,Veledina2011,Veledina2013}. In MAXI J1820+070, during the rising hard state, i.e., before transitioning to the soft state, the illumination of the hard X-ray Comptonization emission on the disk is weakening according to the broad X-ray spectral fits \citep{you2021}, as the hard X-ray Comptonization is contracting towards the BH \citep{kara2019,DeMarco2021}. It was hypothesized that the source of X-ray Comptonization may act as an outflowing corona during the rising hard state of MAXI J1820+070, contributing to the observed decrease in its illumination \citep{you2021}. Discriminating among different physical scenarios for the hard state requires strong observational constraints. Both spectral and timing properties should be considered\citep{Uttley2014}.

\subsection{Measurements and nature of variability }

In addition to the multi-wavelength observation on a long timescale of days, as mentioned above, the fast variability on a timescale of seconds in the inter-bands is also crucial in revealing the physics of the X-ray emission \citep{vanderKlis2006,Uttley2014,Belloni2014,Bollemeijer2025,Bollemeijer2025b,gao2025,Xu2025,Zhan2025} and the accretion–ejection connection \citep{Tetarenko2021,Paice2021,Vincentelli2021,Zhang2022,Bu2021}.  
One can construct X-ray variability spectra by calculating the PDS for each energy channel and integrating it over a specific frequency range to obtain the variance in that channel. This yields the root-mean-square (RMS) spectrum, which characterizes the variability of spectral components within that frequency range \citep{vanderKlis2006,Vaughan2003}. \cite{Wilkinson2009} introduced the technique of covariance spectra, which shares similarities with RMS spectra but is more robust in low signal-to-noise regimes. It measures the covariance between a given energy channel and a broader reference band, enabling the isolation of variability contributions from different spectral components on distinct timescales. Covariance spectra and ratios derived from XMM-Newton observations of GX 339-4 revealed enhanced disk blackbody variability relative to the Comptonized emission below one keV on timescales longer than one second\citep{Wilkinson2009, Uttley2011}, which are caused by instabilities in the disc itself. As a result, we aim to extend the covariance spectrum into the high-energy band, enabling us to study X-ray Comptonization from a timing perspective. It is essential to note that the power-law Comptonization spectrum decreases at higher energies, with a cutoff that reaches approximately 100 keV. This cutoff is typical for the hard spectral states observed in BHXRBs \citep{Gierlinski1997,Zdziarski2004,You2023ApJ}. Therefore, it is crucial to ensure that X-ray observations extend into the high-energy band.


\subsection{Detailing properties of MAXI J1820+070}

In 2018, the outburst of MAXI J1820+070 was discovered in X-rays by the Monitor of All-sky X-ray Image (MAXI) on March 11\citep{Kawamuro2018}.
The long-term and high-frequency observation of MAXI J1820$+$070 by the \emph{Insight}-HXMT took place from March to October 2018\citep{you2021}. During this period, \emph{Insight}-HXMT monitored the outburst over 140 times, accumulating a total exposure of 2560 ks. The high statistics for high-energy photons and the broad energy range (1–250 keV) of \emph{Insight}-HXMT enable detailed timing analysis of high-energy and broad-band variability\citep{kara2019,Buisson2019,DeMarco2021}. 
In this study, we focus on the variability properties of broadband noise (BBN).
We utilize data from \emph{Insight}-HXMT to investigate the broadband X-ray variability of the black hole X-ray binary MAXI J1820+070 during its hard state\citep{Zdziarski_2021a,Stiele_2020}. By applying covariance spectral analysis to the variability over different timescales, we explore the behavior and evolution of BBN across a wide energy range. Our objective is to probe the possible presence of multiple Comptonization regions\citep{Ingram2012,Veledina2013,Zdziarski_2021a}. More details on the data used and on the data analysis are given in Sect. \ref{sec:data}. In Sect. \ref{sec:res}, we present the broadband coherence and covariance ratios on different timescales, and characterize the observed variability through modeling of the covariance spectra. In Sect. \ref{sec:dis}, to interpret the observed variability properties, we explore the possible origins of the hard X-ray emission.

\section{Observations and Data Reduction} \label{sec:data}

The X-ray outburst of MAXI J1820+070 began with a rapid increase in flux, peaking around MJD 58200. This was followed by a gradual decline that lasted until approximately MJD 58290. Subsequently, the source underwent a re-brightening between MJD 58290 and MJD 58305. The hard X-ray flux then decreased dramatically around MJD 58305, indicating a transition from a hard to a soft state. After remaining in the soft state for more than two months, from MJD 58305 to MJD 58380, the hard X-ray flux began to increase around MJD 58380 as the source transitioned back to the hard state. 
Figure \ref{lc} illustrates the \emph{Insight}-HXMT light curves for this outburst, observed by the low-energy (LE, 1 to 10 keV), medium-energy (ME, 10 to 30 keV), and high-energy (HE, 27 to 150 keV) instruments. 
In this study, we focus on the spectral-timing analysis of hard X-ray emissions from the corona, utilizing over 140 \emph{Insight}-HXMT observations that cover the source during its initial hard state (MJD 58191–58300). Due to the enormity of the dataset, in this paper, we select five representative epochs from the total \emph{Insight}-HXMT observations, as shown in Figure \ref{lc}, to illustrate our results. Before MJD 58230, the HE count rates in the 28-150 keV are above $2\times10^{3} $ cnt/s, providing high-quality timing results. Therefore, all five selected epochs are drawn from this period. Observations after MJD 58230 show qualitatively similar behavior, but with poorer statistical quality due to the lower count rates. The information of the selected epochs is detailed in Table \ref{tab:epochs}.

\begin{table}[htbp]
    \centering
    \caption{Observation epochs}
    \label{tab:epochs}
    \begin{tabular}{cccc}
        \toprule
        Epoch & MJD & QPO frequency (Hz) \\
        \midrule
        1 & 58194.53 & 0.029 \\
        2 & 58203.21 & 0.042 \\
        3 & 58211.98 & 0.067 \\
        4 & 58217.42 & 0.081 \\
        5 & 58228.62 & 0.130 \\
        \bottomrule
    \end{tabular}
\end{table}



We extract the time-averaged energy spectra and light curves from three instruments using the \emph{Insight}-HXMT Data Analysis software (HXMTDAS) v2.05. The light curves are not background-subtracted.
To study the energy-dependent timing properties, we extract the PDS from 50 energy channels of interest within 2-150 keV in total: 20 logarithmically spaced energy bands within the 2-10 keV energy range for LE, 10 logarithmically spaced energy bands within the 10-21 keV energy range for ME, 20 logarithmically spaced energy bands within the 30-150 keV energy range for HE. The 21--24 keV data are excluded due to the photoelectric effect of electrons in Silver K-shell in the ME detectors. The data in the 24-30 keV range are also excluded due to the poor signal-to-noise ratio.

\section{Data Analysis and Results} \label{sec:res}

\subsection{X-ray timing analysis}

\begin{figure}
\centering
\includegraphics[width=0.45\textwidth]{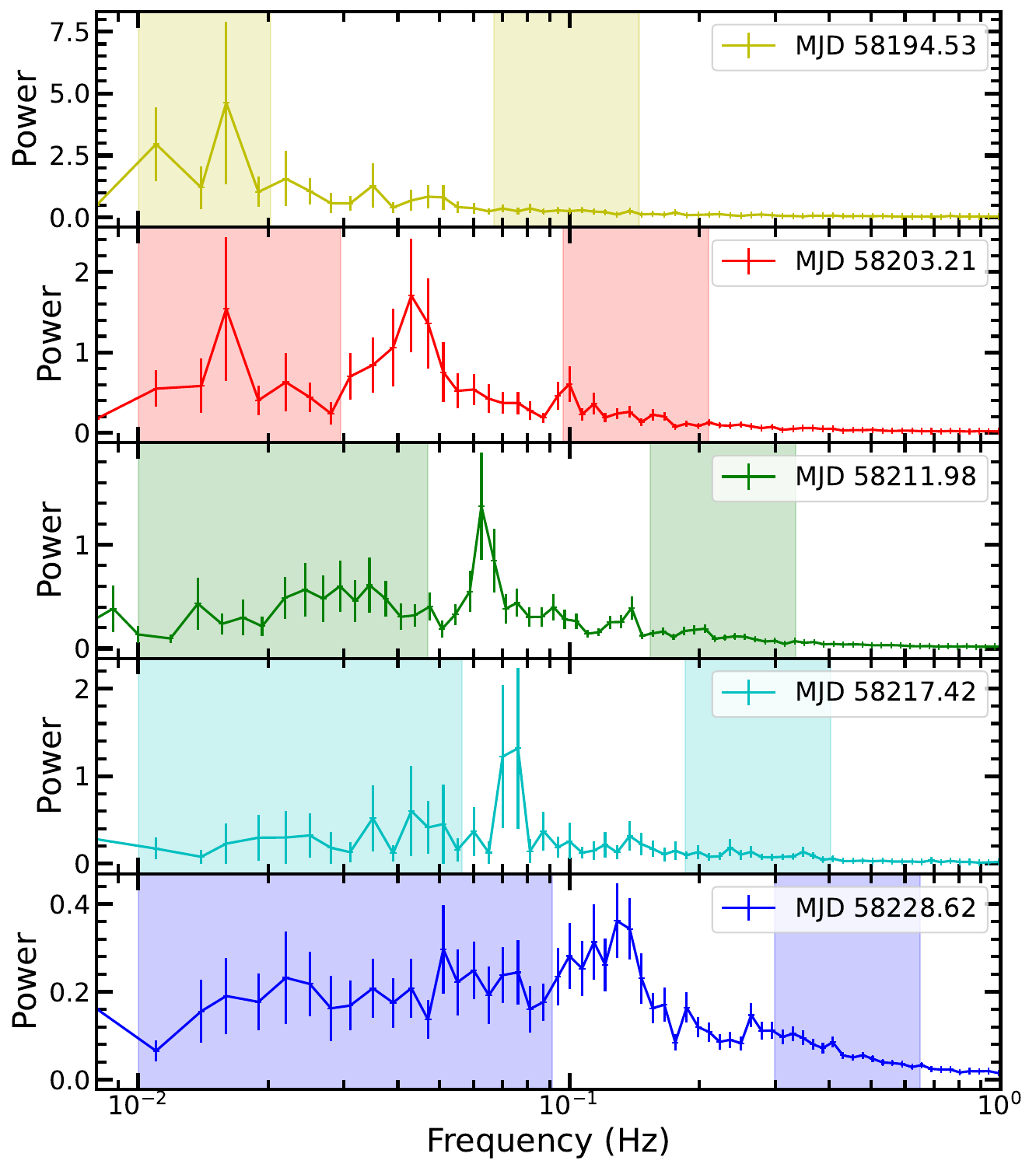}
\caption{
Power density spectrum (PDS) of each epoch in the hard state. The color scheme of the observations is consistent with that shown in Figure \ref{lc}. The PDS are derived from the 2–10 keV light curves obtained with the LE detector of \emph{Insight}-HXMT. The centroid QPO frequencies for the five epochs are 0.029, 0.042, 0.067, 0.081, 0.130 Hz, respectively. The shaded regions denote the frequency ranges corresponding to the long- and short-timescale broadband noise in each epoch.
}
\label{pds}
\end{figure}

\subsubsection{Power density spectra}

We utilize the software {\it Stingray} for timing analysis of X-ray data \citep{Huppenkothen2019} to calculate the PDS. 
The lightcurve is divided into continuous intervals of 100 seconds, with a time resolution of 0.01 seconds, which corresponds to a Nyquist frequency $f_{\text{Nyq}} = 50$ Hz. Then, we compute periodograms for each interval and average them across observations. The PDS is normalized using rms-squared normalization, specifically to the fractional RMS \citep{Miyamoto1991}. These PDS reveal significant variability across a wide frequency range observed by \emph{Insight}-HXMT, extending from soft to hard X-rays.


The Poisson statistics of the power spectrum, denoted as \( P_{\text{noise}} \), is calculated using the formula \( P_{\text{noise}} = \langle \Delta x^{2} \rangle / (\langle x \rangle^{2}) f_{\text{Nyq}} \), where the variable x represents the photon count rate. \( \langle \Delta x^{2} \rangle \) represents the average of the squared error bars of the light curve. 
For the observations made by \emph{Insight}-HXMT, the dead time \( \tau_{d} \) is approximately 20 µs for both HE and LE bands, and 250 µs for the ME band. The frequency range typically analyzed in stellar-mass BH systems is well below \( 1/\tau_{d} \). Given the minimal dead time for HE and LE, our timing results will not be affected by the deadtime. 
For the ME band, the deficit in the power density due to deadtime is approximately 1.9\%, which is less than the background uncertainty of about 3\%. Therefore, the effect of dead time can be reliably ignored \citep{Ma2021}. The resulting PDS of the five selected epochs, after subtracting Poisson noise, are shown in Figure \ref{pds}.

During the hard state from MJD 58191 to MJD 58300, a low-frequency quasi-periodic oscillation (QPO) was detected in 76 observations of \emph{Insight}-HXMT \citep{Ma2021}. The QPO centroid frequency \( f \) varied between 0.03 and 0.7 Hz. The quality factor \( Q \), defined as the ratio of the QPO centroid frequency \( f \) to its full width at half maximum \( \Delta f \), was determined to be greater than 2.5 \citep{Ma2021}.
In this work, our primary goal is to investigate the BBN properties of the source. For this reason, we first excluded variability associated with both the QPO and its harmonic. For each observation in which a QPO with centroid frequency \( f \) is detected, we selected two frequency ranges: \( 0.01-0.7f \) Hz, referred to as {\it long timescale}, and \( 2.3f-5.0f \) Hz, referred to as {\it short timescale}. Based on the measured FWHM of the QPO, these two frequency bands are expected to be dominated by BBN. They correspond to the flat-top and declining portions of the PDS, respectively.
Between MJD 58192 and MJD 58197, no QPO signals were detected. To maintain consistency with the other data, we applied the same long- and short-timescale frequency ranges to the data collected during this stage as used for the observation on MJD 58197 (obsid: P0114661005). 

\begin{figure*}
\centering
\includegraphics[width=\textwidth]{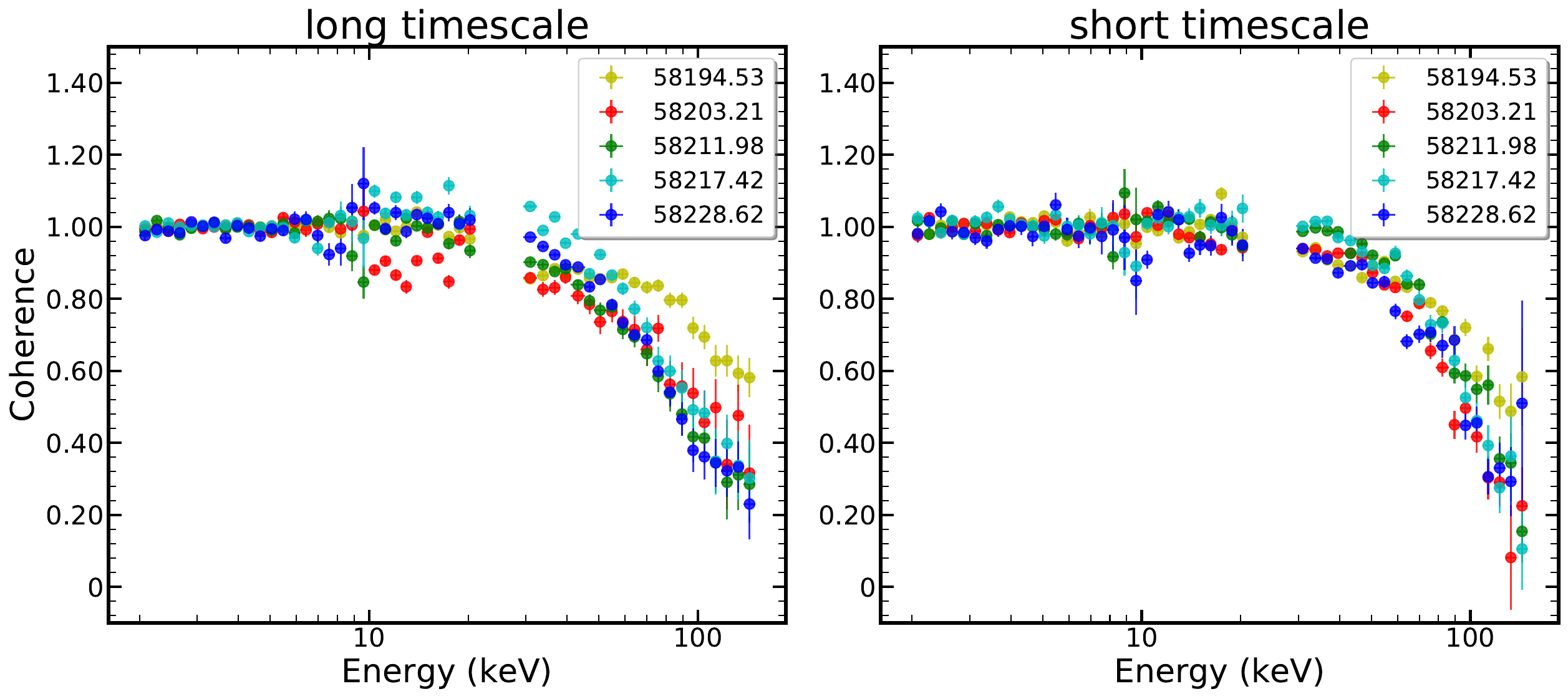}
\caption{
Coherence as a function of energy over the long timescale (left panel) and short timescale (right panel) frequency ranges for our five observation epochs. The coherence spectra are derived from the five epochs of \emph{Insight}-HXMT data, including the rise and decay of the hard state outburst. The 2–10 keV band is selected as the reference.
}
\label{coh}
\end{figure*}

\subsubsection{Energy-dependent coherence}

The derived PDS across soft and hard X-ray energies show that both the magnitude and shape of the noise component change significantly with energy \citep{Yang2022}. To further examine the energy dependence of the BBN variability, we first measure the intrinsic coherence \(\gamma^{2}\) as a function of energy \citep{Vaughan_1997, Uttley2014}, using the 2-10 keV band as the reference band. For each light curve across the 50 energy channels, we calculate the intrinsic coherence between it and the reference light curve on both long and short timescales, following the procedure outlined in \cite{Uttley2014}. For further details, please refer to the Appendix.

The coherence function quantifies the degree of linear correlation between two light curves within a specified frequency range \citep{Vaughan_1997, Uttley2014}. 
The intrinsic coherence as a function of energy, derived on both long and short timescales for the selected epochs, is shown in Figure \ref{coh}.
We find that the coherence changes with energy. On long timescales, as shown in the left panel of Figure \ref{coh}, the coherence remains stable around unity in the 2-10 keV range, as expected since this band is used as the reference. In the 10–30 keV range, the variability also remains coherent, indicating that the soft X-ray emission varies in a similar linear manner across these energies. Interestingly, the coherence begins to decline with energy in hard X-ray bands ($> 30$keV). This behavior suggests that the hard X-ray variability above $\sim30$ keV becomes increasingly incoherent with respect to the soft X-ray variability in the 2–10 keV band. 
On short timescales, a similar trend is exhibited, as illustrated in the right panel of Figure \ref{coh}. Note that the incoherence is also observed in other epochs of \emph{Insight}-HXMT data throughout the hard state on both long and short timescales. 
The evolution of intrinsic coherence with energy suggests the presence of at least two distinct variability components in the X-ray band. One component varies linearly with the soft band (2-10 keV) component, dominating at tens of keV, while the other varies \textit{incoherently} with the soft component, dominating at higher energies on both timescales.



\subsubsection{The root mean square and covariance spectra}

The time-averaged X-ray spectra alone do not capture how the individual spectral components vary relative to each other on timescales of seconds to minutes. Instead, the PDS and coherence analyses reveal significant differences in variability across various energies and frequencies \citep{Paice2021,Tetarenko2019,Tetarenko2021}. To further investigate the properties of the variability and their correlations, energy-dependent spectral-timing analyses are essential.



\cite{Wilkinson2009} developed a technique known as the covariance spectrum. This covariance technique can be considered a matched filter applied to the data. By utilizing the PDS of both the reference band and the band of interest, we can derive the cross-spectrum and subsequently obtain a Fourier frequency-resolved covariance spectrum. This approach allows us to identify the variability in the energy channel being examined that is correlated with the reference band \citep{Uttley2014}.
The covariance spectrum is closely related to the RMS spectrum. The covariance can be expressed as:
\begin{equation}
C v\left(\nu_{j}\right) = \langle x\rangle \sqrt{\gamma^{2} \left(\nu_{j}\right)} \sqrt{\left(\bar{P}_{X}\left(\nu_{j}\right) - P_{X, \text{noise}}\right) \Delta \nu_{j}},
\end{equation}
where $\gamma^{2}$ represents the coherence between two light curves within the specified frequency range. The term $\sigma_{\rm X }(\nu_{j}) \equiv \sqrt{\left(\bar{P}_{X}\left(\nu_{j}\right) - P_{X, \text{noise}}\right) \Delta \nu_{j}} $ indicates the fractional RMS of the energy channel within that frequency range. It is clear that when the coherence is unity, the covariance and RMS are identical. 
For further details, please refer to the Appendix.

We can obtain the RMS spectrum by integrating the PDS over a specified frequency range. The RMS is background-corrected using the formula \( RMS = \sqrt{P} (S + B)/S \), where S and B represent the source and background count rates, respectively, and P is the power calculated over the selected frequency \citep{Bu2015}. The RMS spectrum illustrates the overall intrinsic variability, while the covariance spectrum emphasizes the variability that correlates with the reference band. If there is a single variability pattern across the entire energy band and coherence is perfect (unity), the shape of the covariance spectrum will match that of the RMS spectrum \citep{Wilkinson2009}.

\begin{figure}
\centering
\includegraphics[width=0.45\textwidth]{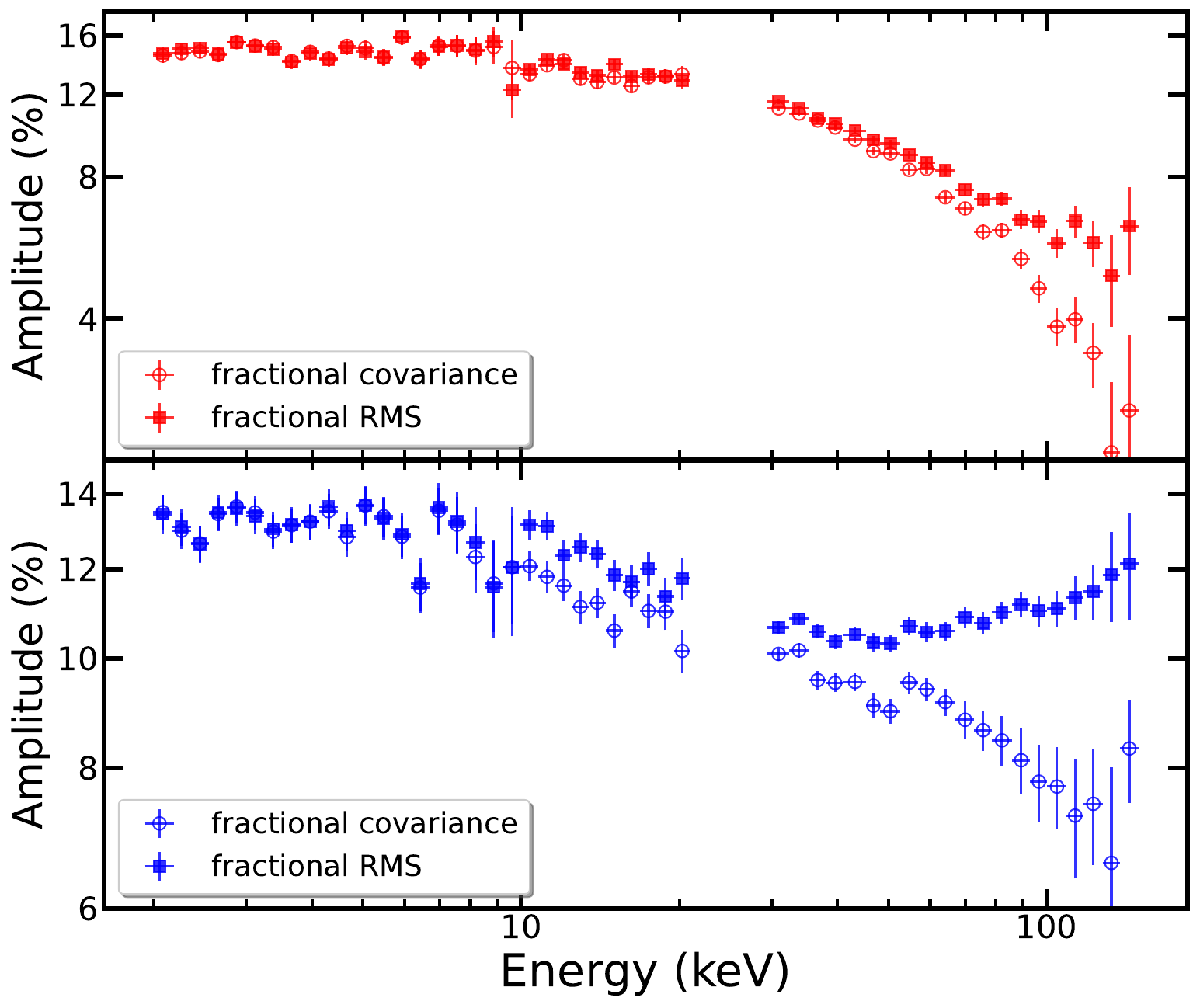}
\caption{
Fractional covariance and RMS spectra. The top panel shows the fractional covariance (red circles) and fractional RMS (red squares) spectra on the short timescale. The bottom panel shows the corresponding spectra on the long timescale (blue circles and blue squares, respectively). The 2–10 keV band is selected as the reference.
}
\label{fcov}
\end{figure}

\subsubsection{Fractional covariance spectra}

By dividing the covariance spectra by the time-averaged spectra, we derive the fractional covariance spectra. Similar to the fractional RMS spectra, the fractional covariance spectra provide insight into the relative contributions of different components to the observed variability.\citep{Cassatella2012}. 

Figure \ref{fcov} displays the fractional covariance and RMS spectra obtained on both long and short timescales, using the 2-10 keV reference band. At low energies, the fractional RMS spectra and the fractional covariance spectra closely align, appearing almost flat. And the coherence approaches unity. In contrast, at high energies, the fractional covariance spectra exhibit a power-law slope. Notably, the fractional covariance spectrum over the long timescale is significantly harder than that over the short timescale.



\subsubsection{Covariance ratio}

\begin{figure}
\centering
\includegraphics[width=0.47\textwidth]{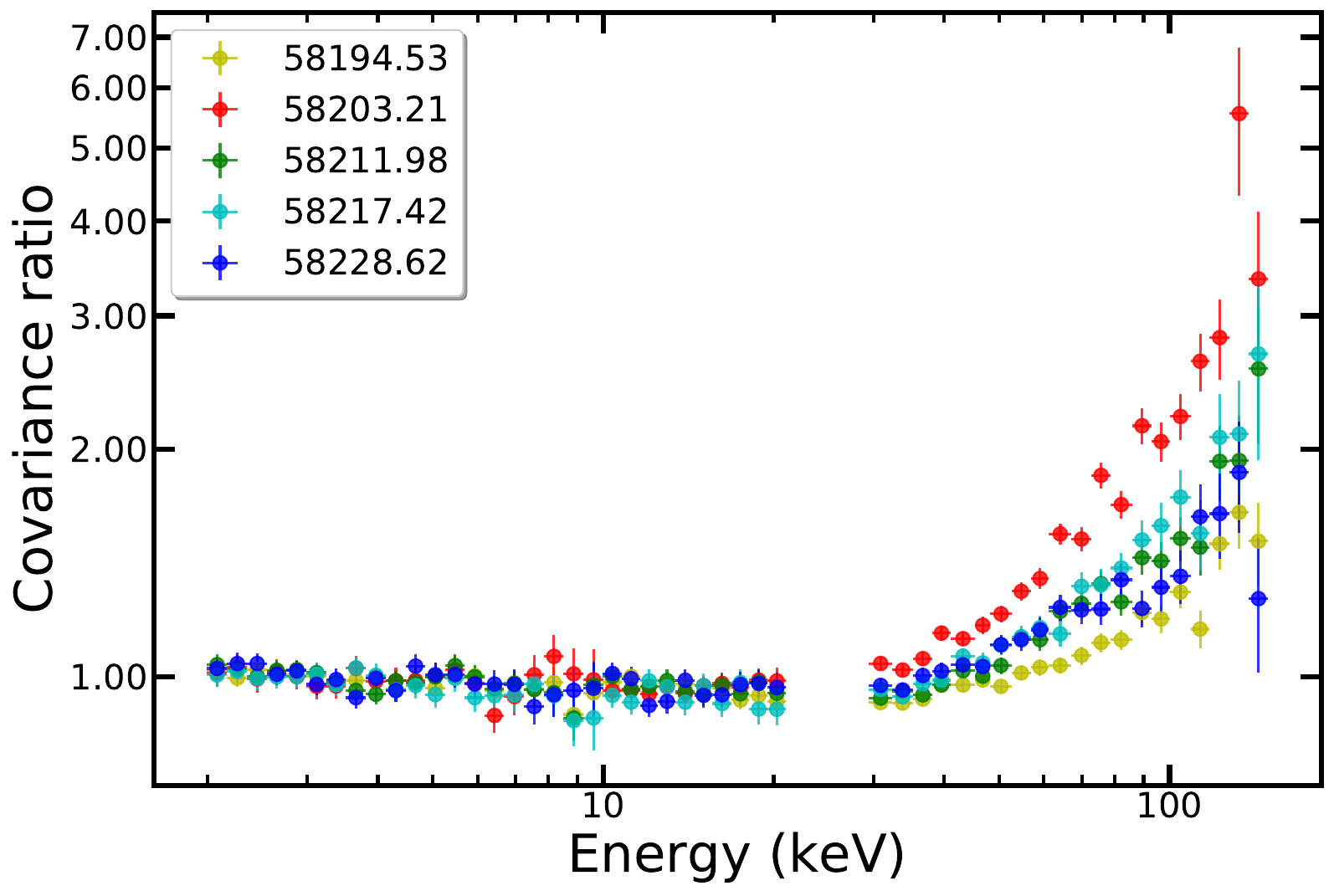}
\caption{
The covariance ratio of the five representative epochs of the \emph{Insight}-HXMT observations during the hard state. The reference band is 2–10 keV. To highlight the differences at high energy, the ratios at different times have been rescaled by multiplying factors of 1.20, 0.67, 0.83, 0.89, and 0.94, respectively. The legend on the top left of the panel indicates the observation time of the covariance ratio in MJD.} 
\label{cov_ratio}
\end{figure}

We calculate covariance ratios by dividing the long-timescale covariance spectrum by the short-timescale covariance spectrum. By utilizing more than 70 observations from \emph{Insight}-HXMT during the X-ray hard state, we investigated the evolution of the covariance ratio throughout the outburst. As mentioned earlier, due to the enormity of the dataset, here we only display the covariance ratios for the five representative epochs of observations in Figure \ref{cov_ratio}.
The covariance ratios remain relatively constant with energy below approximately 30 keV. However, above 30 keV, the covariance ratios increase with energy, indicating a \textit{\ {hard excess}} of variability on long timescales. In other words, the trend of coherence toward higher energies on long timescales differs from that on short timescales. This suggests that the variability in the hard band exhibits distinct behaviors on both long and short timescales. By analyzing the significant changes in the covariance ratio with energy, we can see in a model-independent way that the variability of the hard band ($>$ 30 keV) emission, which is linearly correlated with the variability of the soft band (2-10 keV) emission, exhibits distinct behaviors on both long and short timescales. An increase in the covariance ratio may indicate additional Comptonization variability on long timescales.


A clear evolution in the covariance ratio is apparent across different epochs (Figure \ref{cov_ratio}). In the first epoch (MJD 58194), the covariance ratio increases gradually with energy but displays a relatively shallow slope. In subsequent epochs, this energy-dependent trend becomes more pronounced. On MJD 58203, near the peak of the long-term light curve, the covariance ratio exhibits a significantly steeper slope than in other observations. After MJD 58203, the slope gradually decreases. This evolution indicates distinct X-ray variability behavior during the rising phase (before MJD 58200) and the decaying phase (after MJD 58200).


\subsection{Modeling of the covariance spectra}

\begin{figure*}
\centering
\includegraphics[width=0.95\textwidth]{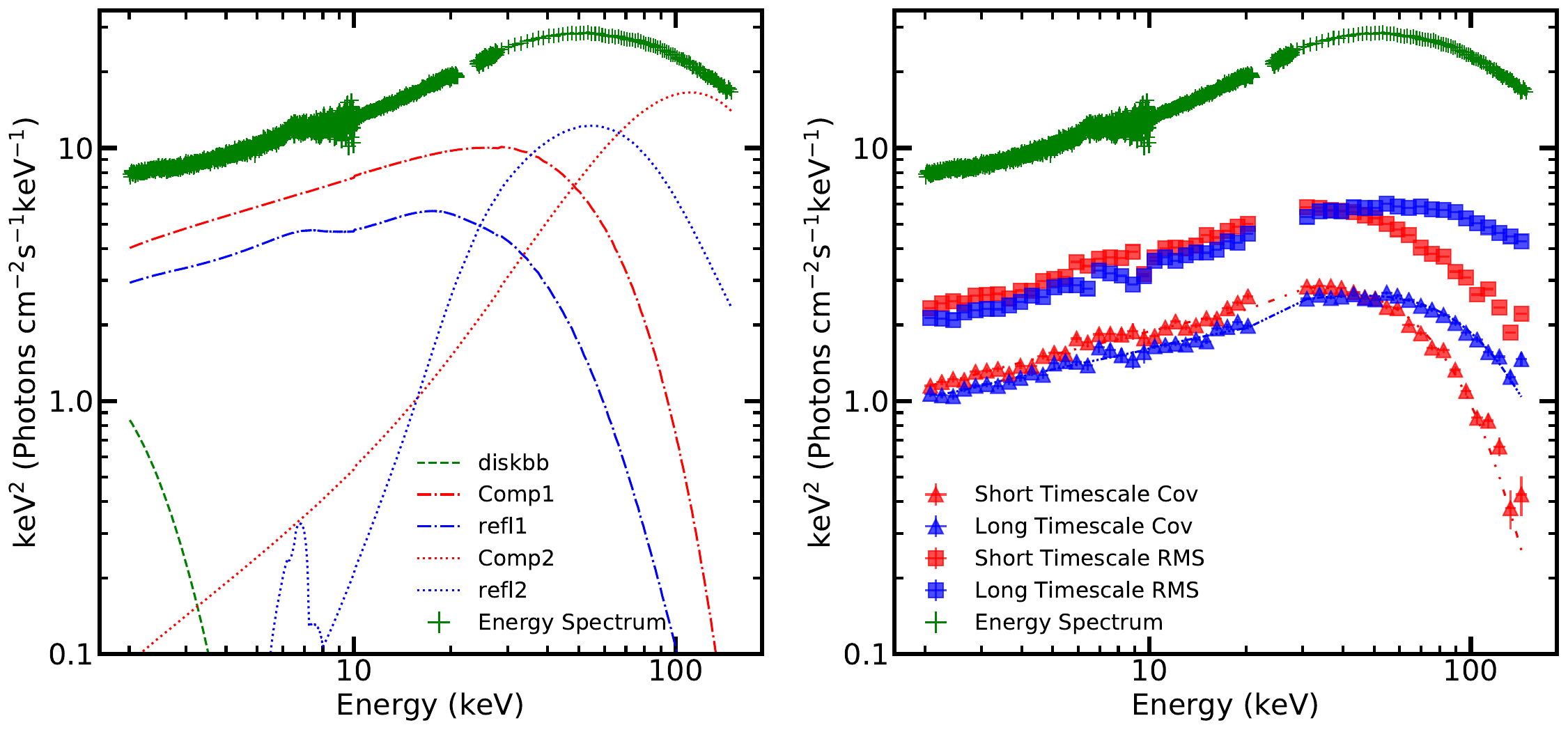}
\caption{
Spectral fitting of the time-averaged energy spectrum and covariance spectra of epoch 3, in the range of 2-150 keV. The left panel: the unfolded time-averaged energy spectrum (green), fitted with the model {\tt Tbabs*(diskbb+$relxillCp_{1}$+$relxillCp_{2}$)}. The softer and harder Comptonization components are shown by the red dot-dashes and red dots, respectively. The corresponding reflection components are shown by the blue dot-dashes and blue dots, respectively. The right panel: The energy spectrum (green), RMS spectra (squares) and covariance spectra (triangles). The covariance spectra are fitted with model {\tt Tbabs*(diskbb+Nthcomp)}. The blue and red dashed curves represent the best-fitting models of the covariance spectra on the long and short timescales, respectively. The RMS spectra have been shifted up by multiplying by a factor of 2.
}
\label{spec_covs}
\end{figure*}

In this study, we examined the covariance spectra, which are related to spectral variability. We have observed a significant hard excess of the variability over long timescales. To identify the origin of the hard excess, in this section, we try to fit the long- and short-timescale covariance spectra.

\cite{Zdziarski_2021a} found that the spectra during the hard state necessitate using two different Comptonization components and two corresponding reflection components. Following this work, we conduct spectral fits using data from \emph{Insight}-HXMT in the 2–150 keV energy range with {\tt XSPEC}. And, the time-averaged spectral model used for fitting is {\tt Tbabs*(diskbb+$relxillCp_{1}$+$relxillCp_{2}$)}.

For the covariance spectra, we employed a simple model {\tt Tbabs*(diskbb+Nthcomp)}, as the reflection features are not required due to the limited energy resolution of the covariance spectra. For each observation, we simultaneously fit the short- and long-timescale covariance spectra while constraining the absorbing column density of {\tt Tbabs} to $N_\mathrm{H} = 1.5\times 10^{21} \rm cm^{-2}$, which corresponds to the value used in the fits of the time-averaged energy spectra \citep{you2021}. The remaining parameters were allowed to vary freely between the covariance spectra. 72 of 76 covariance spectra can be fitted well with the reduced $\chi^{2} < 1.0$.

The best-fitting covariance spectra and the time-averaged spectrum are presented in Figure \ref{spec_covs}. Part of the fitted parameters for energy and covariance spectra are listed in Table \ref{tab:param}. The covariance spectra on different timescales are consistent at low energies but show different feature on at high energies, which implies that the hard excess is not associated with the soft component, but instead arises entirely from the hard X-ray emission.

Assuming that the variability of the Compton component mainly results from changes in its normalization, the covariance spectrum in its dominant energy band would closely mirror the shape of the time-averaged energy spectrum. 
The best-fitting parameters for the short-timescale covariance spectrum are generally consistent with those from the long-timescale covariance spectrum, except for the electron temperature. Interestingly, the electron temperature measured on the long timescale is significantly higher than that measured over the short timescale.

\begin{table*}[htbp!]
\centering
\caption{Fit parameters for energy and covariance spectra}
\begin{tabular}{l|c|c|c|c}
\hline
\textbf{Parameter} & \textbf{Mean($relxillCp_{1}$)} & \textbf{Mean($relxillCp_{2}$)} & \textbf{long timescale cov} & \textbf{short timescale cov} \\
\hline
$\Gamma$ & $1.68 \pm 0.01$ & $1.11 \pm 0.02$ & $1.74 \pm 0.02$ & $1.72 \pm 0.04$  \\
$kT_{e}$ & $11 \pm 1$ & $28 \pm 1$ & $32 \pm 4$ & $18 \pm 1$ \\
$N_{th}$ & - & - & $0.87^{+0.06}_{-0.5}$& $0.66 \pm 0.24$  \\
$R_{th}$ & 0.9 & 2.0 & - & -  \\
\hline
\end{tabular} \\
\noindent \textbf{Note.} The best-fitting results for the time-averaged energy spectrum using the model {\tt Tbabs*(diskbb+$relxillCp_{1}$+$relxillCp_{2}$)} and for long/short-timescale covariance spectra {\tt Tbabs*(diskbb+Nthcomp)}. $\Gamma$ is the power-law index of the Compton component, $kT_{e}$ is the electron temperature in units of keV, $N_{th}$ is the flux density of a Compton component at 1 keV, and $R_{th}$ is the reflection fraction. 
\label{tab:param}
\end{table*}

\begin{figure}
\centering
\includegraphics[width=0.45\textwidth]{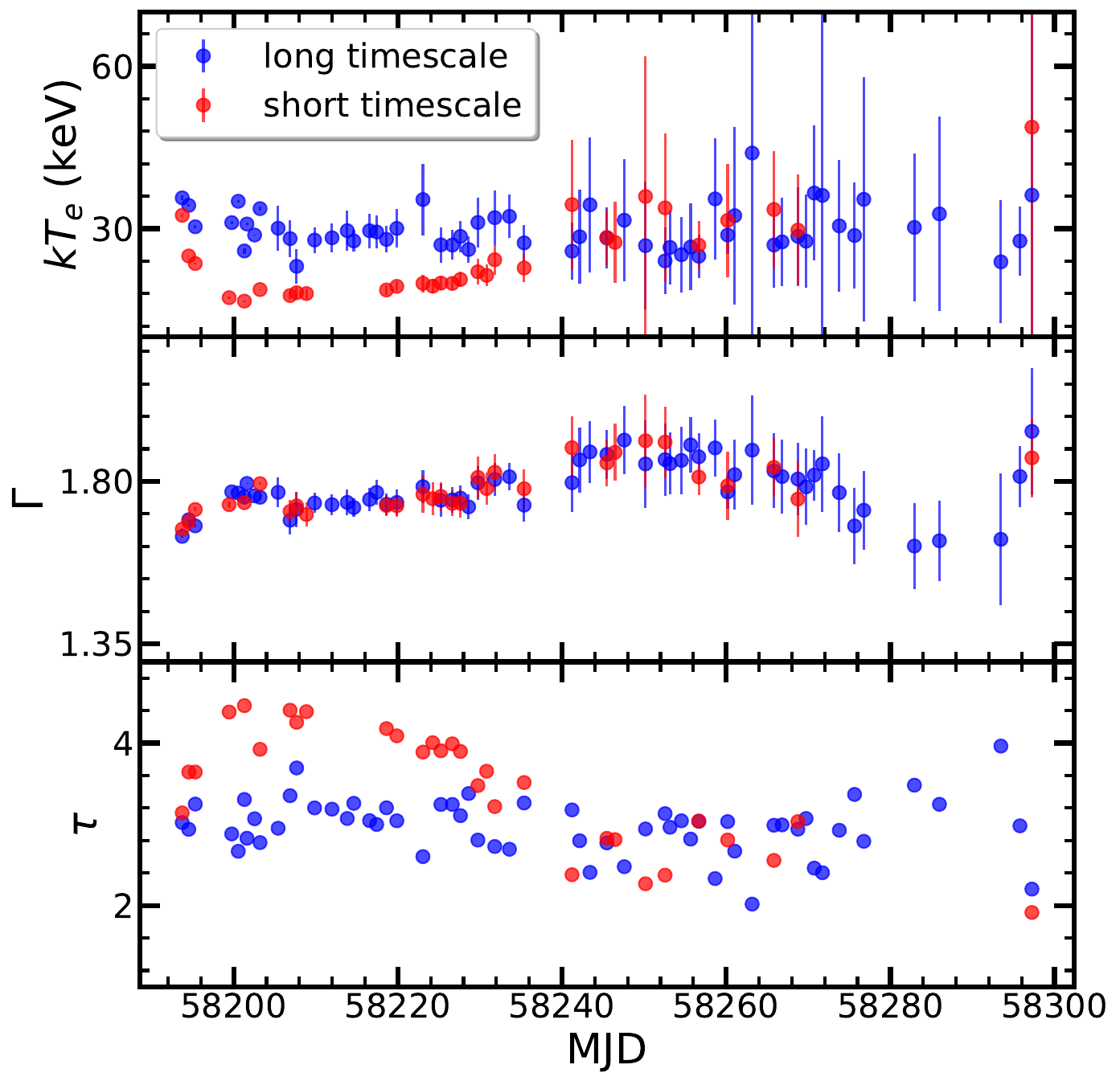}
\caption{Temporal evolution of the best-fitting parameters of the covariance spectra on long (blue dots) and short (red dots) timescales. From top to bottom, the panels show the evolution of the electron temperature (in keV), the photon index $\Gamma$, and the optical depth $\tau$ of the corona, respectively.}
\label{cutoffe}
\end{figure}

We examine the temporal evolution of the spectral fit parameters. During the early hard state, the long-timescale electron temperature exceeds the short-timescale value (Figure \ref{cutoffe}). The long-timescale temperature remains relatively constant throughout the hard state, while the short-timescale temperature first decreases then rises, displaying an anticorrelation with X-ray luminosity. The photon index remains consistent between both variability components throughout the hard state. We estimate the optical depth of the Comptonizing component using Equation (2) from \cite{Yan2020}. The optical depth exhibits markedly different temporal evolution on short versus long timescales (Figure \ref{cutoffe}).

\section{Discussion} \label{sec:dis}




In this study, we study energy-dependent coherence up to 150 keV and discover that the hard X-ray variability above approximately 30 keV becomes increasingly incoherent with the soft X-ray variability (2-10 keV) (see Fig. \ref{coh}). This finding suggests that at least two distinct Comptonization components contribute to the overall X-ray variability.
One component varies linearly with the soft X-ray variability and primarily emits below 30 keV. The other component varies incoherently with the soft X-ray variability and primarily emits above 30 keV. Hereafter, we refer to these components as \textbf{the coherent component} and \textbf{the incoherent component}, respectively.

We then employ the energy-dependent coherence to select the coherence component across different Fourier frequency ranges. From this analysis, we derive the covariance spectra for both long- and short-time scales. By calculating the ratio of the covariance spectra between these two timescales, we uncover a low-frequency excess at high energies (see Fig. \ref{fcov}).

Furthermore, we fit the covariance spectra for both timescales, using the model {\tt Tbabs*(diskbb+Nthcomp)}, to quantitatively characterize the covariance spectra (see Fig. \ref{cov_ratio}).
The parameters obtained from the short-timescale covariance spectrum are generally consistent with those from the long-timescale covariance spectrum, except for the electron temperature. Specifically, the electron temperature measured on the long timescale is significantly higher than that measured over the short timescale.
Moreover, the evolution of electron temperature differs notably between the two timescales (see Fig. \ref{cutoffe}). The electron temperature associated with long-timescale variability is higher than that associated with short-timescale variability, indicating a deficit in hard X-ray variability on the short timescale. Specifically, the electron temperature for the long-timescale covariance spectrum remains relatively constant throughout the hard state, while the short-timescale electron temperature initially decreases before exhibiting a distinct increase. 
Thus, the evolution of the electron temperature for short timescale variability directly influences the evolution of the covariance ratio. In this case, we suggests that the increase in covariance ratios above approximately 30 keV, corresponding to the hard excess on long timescales, is caused by a deficit of hard X-ray variability on short timescales.


\subsection{On the incoherent variability: its origin}

The variability of Comptonization emission is influenced by variability of both the seed photons and electrons. Thus, the observed incoherent variability is likely a consequence of the complexity of these two components. In this section, we will discuss the potential regions where Comptonization occurs and the origins of the seed photons in order to better understand the nature of this variability. Furthermore, we will explore the association between the Comptonizing source and the flows near the BH.

\subsubsection{Multiple Comptonization emission}

One possibility accounting for the observed incoherent variability is that multiple Comptonization emissions, as proposed in previous studies, are involved. For instance, \cite{Dzielak2021} studied the BBN of the PDS for MAXI J1820+070. They found that the BBN could be modeled with four Lorentzian components, each exhibiting significant variability. At least two Comptonization components, characterized by different electron temperatures and optical depths, are required to accurately fit both the variability and time-averaged spectra. \cite{Zdziarski_2021a} studied the broadband spectra and found that a multi-Comptonization model can fit the spectra well. In their scenario, the system consists of two distinct Comptonization components: a harder component originating from a larger-scale height accretion flow located downstream of the truncation radius, and a softer component forming a corona over the inner disk. These separate regions produce distinct Compton reflection signatures. \cite{Wang2024} examined the evolution of the frequency-dependent RMS-flux relation in MAXI J1820+070 using \emph{Insight}-HXMT data. The timing features and their evolution support the existence of two distinct Comptonization regions: one that remains radially extended, allowing fluctuations to propagate from the disk to the corona, and another that is vertically extended and continues to expand during the later stages of the initial hard state. 


It should be noted that we cannot completely rule out a single source of Comptonization. The exponential cut-off function is highly nonlinear\citep{Uttley2024}, such that variations in the cut-off energy could also lead to a reduction in coherence. 
If the variability of the Comptonizing component arises not only from changes in its normalization but also from variations in the cut-off energy, the resulting variability spectrum may differ from the time-averaged spectrum. Owing to the non-linear nature of the cut-off function, such spectral variability can also lead to a decrease in coherence. It seems plausible that the incoherent variability could be partially attributed to this type of non-linear spectral variability. However, to draw more definitive conclusions, we would need further quantitative modeling that might suggest the presence of distinctly varying components beyond mere non-linearity. In this work, we assume that the variability of the Comptonizing component is primarily driven by changes in its normalization.




\subsubsection{Multiple fields of seed photon}



Another possibility responsible for the incoherent variability is associated with the complexity of the seed photon field. To examine this, we first consider disk photons as the seed photons involved in the inverse Compton scattering process. \cite{Uttley2024} constructs a model in which the interplay between the disk seed photons and the corona is taken into account to simulate the Comptonization variability.
Based on their model, we conducted a simulation considering 10 groups of disk photons with different seed energies. All of these photons were Compton up-scattered by thermal electrons in the corona via inverse-Compton processes. Then we derive the overall variability spectrum over a broad energy band. Since the BBN is thought to arise from propagating accretion rate fluctuations, the input seed photon variations of all 10 groups are expected to be almost identical. In this case, we found that we could not reproduce the reduction of the coherence in the hard X-ray band (see Fig. \ref{coh}), as all the resulting power-law emission variations remained coherent. However, when the input seed photon variations of the 10 groups are assumed to be independent of each other, a broad reduction in coherence naturally emerges. This suggests that the seed photons may originate from multiple components, rather than exclusively from the accretion disk.

In this framework, we conclude that the seed photons for inverse-Compton scattering should have two origins: the disk blackbody and synchrotron emission \citep{yang2015}, and it is the Comptonization of synchrotron emission as seed photons that is responsible for the incoherent variability.


\subsection{On the coherent variability: electron temperature on different timescales}

Variability on timescales ranging from a few tens of seconds down to a few tens of milliseconds can arise from all radii, and subsequently propagate inward toward the black hole through the accretion flow \citep{Lyubarskii1997,Kotov2001}. At each radius, fluctuations are expected to propagate most efficiently at frequencies comparable to the local viscous timescale. Consequently, variability generated in the inner regions of the flow occurs at higher frequencies than that arising from larger radii. The fluctuation at each
radius should be viewed as the superposition of the local fluctuation and those propagated from the outer region.
This propagating fluctuation model naturally explains the observed correlations between variability in the soft and hard X-ray bands \citep{Kotov2001,Zhan2025}.

In this work, the spectra of the coherent variability were well fitted with the model {\tt Tbabs*(diskbb+Nthcomp)}. It suggests that the BBN variability, on both long and short timescales, is not dominated by a single spectral component but rather by a combination of a disk and a Comptonization component. The modeling of the covariance spectra aligns with our interpretation in the previous section, indicating that the coherent variability is predominantly attributed to Comptonization of disk seed photons. Within the propagating-fluctuation model, the short-timescale variability arises from the inner region of the accretion flow, while long-timescale variability originates from larger radii \citep{Kotov2001}. 
In this framework, one would expect the Comptonization component of the variability spectra to exhibit progressively higher electron temperatures at higher Fourier frequencies \citep{axelsson2018}.
However, we found that the Comptonization component on short timescales has a lower electron temperature than that on long timescales.
This may indicate that the seed photon associated with Comptonization on the short timescale has a lower temperature than those linked to the long timescale.

To understand the different temperatures of the disk seed photons inferred from the variability spectra, we consider two Comptonization continua.
Specifically, the variability on short timescales originates in a 
central Comptonization region with a large scale height, while the long-timescale variability comes from a region located at somewhat larger radii. \cite{Zdziarski_2021a} suggested that emission from the elevated Comptonization region is reflected off the disk at relatively large radii. 
Similarly, we propose that the Comptonization region with a large scale height primarily interacts with the outer disk at larger radii, whereas the lower Comptonization region interacts with the adjacent inner disk. 
Consequently, the Comptonized disk emission from the elevated region has a lower temperature, leading to the softer variability observed on short timescales.

Within this framework, if the central Comptonization region has a larger vertical extent, it is expected to be illuminated by disk photons originating from larger radii, which have lower temperatures. This results in the Comptonization emission in this region having a lower electron temperature. Therefore, we could interpret the evolution of the electron temperature on different timescales, presented in Fig. \ref{cutoffe}, in terms of structural changes in the central Comptonization region. At the beginning of the hard state, the two Comptonization regions have nearly the same vertical height, making their covariance spectra indistinguishable. During the rise phase of the hard state (MJD 58192 to 58200), the central region extends vertically, increasing its height. As a result, the electron temperature associated with the short-timescale variability decreases, which corresponds to the deficit of hard X-ray variability on the short timescale (see the right panel of Fig. \ref{spec_covs}). During the decay of the hard state (after MJD 58200), as the central region contracts and its spatial extent decreases, the electron temperature increases on short timescales. Ultimately, after roughly MJD 58240, the two regions reach comparable heights, resulting in spectral properties that are indistinguishable once again. 


\section{Acknowledgements}

We thank F.M. Vincentelli, Yi Long, and Zhen Yan for their valuable discussion on spectral analysis. B.Y. is supported by Natural Science Foundation of China (NSFC) grants 12322307, 12361131579, and 12273026; by Xiaomi Foundation / Xiaomi Young Talents Program. The data analysis in this paper have been done on the supercomputing system in the Supercomputing Center of Wuhan University. AJT acknowledges that this research was undertaken thanks to funding from the Canada Research Chairs Program and the support of the Natural Sciences and Engineering Research Council of Canada (NSERC; funding reference number RGPIN-2024-04458). AAZ acknowledges support from the Polish National Science Center grants 2019/35/B/ST9/03944 and 2023/48/Q/ST9/00138. 

\appendix
\subsection{Detailed Calculation of the Coherence and Covariance}

As mentioned in Section \ref{sec:res}, we followed the procedure outlined in \cite{Uttley2014} to derive the coherence and covariance. 

The Fourier cross-spectrum between two light curves x(t) and y(t) with discrete Fourier transforms (DFTs) $\rm X_{n}$ and $\rm Y_{n}$ is defined to be:
\begin{equation}
C_{X Y, n}=X_{n}^{*} Y_{n}.
\end{equation}

In a given frequency bin $\nu_{j}$, the averaged cross-spectrum over M segments and K frequencies per segment should be derived as follows:
\begin{equation}
\bar{C}_{X Y}\left(v_{j}\right)=\frac{1}{K M} \sum_{n=i, i+K-1} \sum_{m=1, M} C_{X Y, n, m}.
\end{equation}
where $\bar{C}_{X Y}\left(v_{j}\right)$ is the estimate of the cross-spectrum from the average of the periodogram in the bin $\nu_{j}$ and $C_{X Y, n, m}$ is the value of a single sample of the periodogram measured from the $m$th segment with a frequency $f_{n}$ that is contained within the frequency bin $\nu_{j}$.

The intrinsic coherence is calculated using the following formula:
\begin{equation}
\gamma^{2}(\nu_{j}) = \frac{|\bar{C}_{XY}(\nu_{j})|^{2} - n^{2}}{(\bar{P}_{X}(\nu_{j}) - P_{X, \text{noise}})(\bar{P}_{Y}(\nu_{j}) - P_{Y, \text{noise}})},
\end{equation}
where \(\bar{C}_{XY}(\nu_{j})\) represents the average cross-spectrum within the specified frequency range. All power spectra and cross-spectra are normalized using Miyamoto normalization, which is adjusted to the fractional RMS according to \cite{Miyamoto1991}.

The $n^{2}$ term accounts for the bias in the cross-spectrum due to the effects of Poisson noise. It is determined by the following equation:
\begin{equation}
\begin{aligned}
  &n^{2}=\left[\left(\bar{P}_{X}\left(\nu_{j}\right)-P_{X, \text { noise }}\right) P_{Y, \text { noise }} 
  \notag\right.
  \\
  \phantom{=\;\;}
  &\left.+\left(\bar{P}_{Y}\left(\nu_{j}\right)-P_{Y, \text { noise }}\right) P_{X, \text { noise }} + P_{X, \text { noise }} P_{Y, \text { noise }}\right] / K M
\end{aligned}
\end{equation}

To prevent autocorrelation, we exclude the channel from the reference band when calculating the intrinsic coherence for the energy channel of interest within the reference band. The PDS is derived from the \emph{Insight}-HXMT light curve without background subtraction. We assume that the background flux remains constant over time, which means the power spectrum of the background light curve is constant across frequencies and does not contribute to the intrinsic coherence \(\gamma^{2}\).

The error on the intrinsic coherence can be estimated as [\cite{Vaughan_1997}, their equation 8]
\begin{equation}
\Delta\gamma^{2}(\nu_{j}) = \gamma^{2}(\nu_{j}) \times \sqrt{ [\frac{2 n^{4} KM}{\left(|\bar{C}_{XY}(\nu_{j})|^{2} - n^{2}\right)^{2}} + \frac{P_{X, \text { noise }}^{4}}{\bar{P}_{X}\left(\nu_{j}\right)^{4}}+\frac{P_{Y, \text { noise }}^{4}}{\bar{P}_{Y}\left(\nu_{j}\right)^{4}}+\frac{KM \delta\gamma^{2}}{\gamma^{4}}]/KM },
\end{equation}
where $\delta\gamma^{2}$ is the statistical uncertainty of the coherence function, $\delta\gamma^{2} = 2\left(1-\gamma^{2}\right)^{2} /(KM\gamma^{2})$.

Following the approach outlined in \cite{Uttley2014}, we derive covariance spectra over both long and short timescales using light curves from 50 specific energy channels, along with a reference band. For this study, we have chosen the 2-10 keV energy range as the reference band. 
For each energy channel of interest, we first construct the power spectra \( P_{X}(\nu_{j}) \) and \( P_{Y}(\nu_{j}) \) from the light curves of the energy channel and the reference band, which we denote as \( x(t) \) and \( y(t) \), respectively. It is important to note that the power spectra discussed here are not noise-subtracted. Using these two light curves, we also compute the cross-spectrum \( C_{XY}(\nu_{j}) \).
To generate the power and cross-spectra, we utilize the Python-based package {\it Stingray}. All power spectra and cross-spectra are normalized using Miyamoto normalization, which is based on the fractional RMS \citep{Miyamoto1991}.
After deriving power spectra and the cross-spectrum, the value of the covariance spectrum in each energy channel of interest is given by:
\begin{equation}
C v\left(\nu_{j}\right) = \langle x\rangle \sqrt{\frac{\Delta \nu\left(\left|\bar{C}_{X Y}\left(\nu_{j}\right)\right|^{2}-n^{2}\right)}{\bar{P}_{Y}\left(\nu_{j}\right)-P_{Y, \text { noise }}}},
\label{cal_cov}
\end{equation}
where the first term $\langle x\rangle$ is the average count in the energy channel of interest and the term \(\Delta \nu_{j}\) indicates the width of this frequency range. The second term is dimensionless; we refer to it as fractional covariance.

Note that the covariance spectrum is closely related to RMS. The equation can be expressed as follows:
\begin{equation}
C v\left(\nu_{j}\right) = \langle x\rangle \sqrt{\gamma^{2} \left(\nu_{j}\right)} \sqrt{\left(\bar{P}_{X}\left(\nu_{j}\right) - P_{X, \text{noise}}\right) \Delta \nu_{j}},
\end{equation}
where $\gamma^{2}$ represents the coherence between two light curves within the specified frequency range. Since the power spectrum utilizes fractional RMS-squared normalization, the term $\sigma_{\rm X }(\nu_{j}) \equiv \sqrt{\left(\bar{P}_{X}\left(\nu_{j}\right) - P_{X, \text{noise}}\right) \Delta \nu_{j}} $ indicates the fractional RMS of the energy channel within that frequency range. It is also essential to apply the background correction to this term, as given by the formula \(RMS = \sqrt{P} (S + B)/S\) \citep{Bu2015}. Additionally, the error on the covariance could be derived via error propagation: 

\begin{equation}
\Delta C v\left(\nu_{j}\right) = \sqrt{ \left( \sqrt{\gamma^{2}(\nu_{j})} \cdot \sigma_{\rm X }(\nu_{j}) \cdot \Delta{\rm \langle x\rangle}\right)^{2} +\left( \dfrac{\sigma_{\rm X}(\nu_{j}) \cdot \langle x\rangle \cdot \Delta\gamma^{2}(\nu_{j})}{2 \sqrt{\gamma^{2}(\nu_{j})}}\right)^{2} + \left( \sqrt{\gamma^{2}(\nu_{j})} \cdot \langle x\rangle \cdot \Delta\sigma_{\rm X}(\nu_{j})\right)^{2} },
\end{equation}
where $\Delta\sigma_{\rm X}(\nu_{j})$ is the error on the fractional rms-spectrum.

\clearpage

\bibliography{reference}{}
\bibliographystyle{aasjournal} 

\end{document}